\documentclass[aps,prl,twocolumn,superscriptaddress,floatfix]{revtex4-1}
\usepackage{makecell}
\usepackage{graphicx}% Include figure files
\usepackage{dcolumn}% Align table columns on decimal point
\usepackage{bm}% bold math
\usepackage{times}% font that prx use
\usepackage{color}
\usepackage{appendix}

\begin{document}
\bibliographystyle{apsrev4-1}
\title{Quantum Oscillations and Coherent Interlayer Transport in a New Topological Dirac Semimetal Candidate: YbMnSb$_2$}

\author{Yi-Yan Wang}
\author{Sheng Xu}
\author{Lin-Lin Sun}
\author{Tian-Long Xia}\email{tlxia@ruc.edu.cn}

\affiliation{Department of Physics, Renmin University of China, Beijing 100872, P. R. China}
\affiliation{Beijing Key Laboratory of Opto-electronic Functional Materials $\&$ Micro-nano Devices, Renmin University of China, Beijing 100872, P. R. China}

\date{\today}

\begin{abstract}
Dirac semimetals (DSMs), which host Dirac fermions and represent new state of quantum matter, have been studied intensively in condensed matter physics. The exploration of new materials with topological states is important in both physics and materials science. In this article, we report the synthesis and the transport properties of high quality single crystals of YbMnSb$_2$. YbMnSb$_2$ is a new compound with metallic behavior. Quantum oscillations, including Shubnikov-de Haas (SdH) oscillation and de Haas-van Alphen (dHvA) type oscillation, have been observed at low temperature and high magnetic field. Small effective masses and nontrivial Berry phase are extracted from the analyses of quantum oscillations, which provide the transport evidence for the possible existence of Dirac fermions in YbMnSb$_2$. The measurements of angular-dependent interlayer magnetoresistance (MR) indicate the interlayer transport is coherent. The Fermi surface (FS) of YbMnSb$_2$ possesses quasi-two dimensional (2D) characteristic as determined by the angular dependence of SdH oscillation frequency. These findings suggest that YbMnSb$_2$ is a new candidate of topological Dirac semimetal.
\end{abstract}
\maketitle

\section{Introduction}

The discoveries of three dimensional (3D) DSMs Na$_3$Bi\cite{PhysRevB.85.195320,liu2014discovery,xiong2015evidence,xiong2016anomalous} and Cd$_3$As$_2$\cite{neupane2014observationCd3As2,YLChen2014stableCd3As2,PhysRevLett.113.027603,liang2015ultrahigh,he2014quantum,li2015negative} have inspired great interest in the study of topological semimetals. In DSMs,  the quasi-particles are Dirac fermions, which are formed by the crossing of fourfold degenerate linear bands and can be described by relativistic Dirac equations\cite{wehling2014dirac}. The breaking of time reversal symmetry or space inversion symmetry will drive DSMs into Weyl semimetals\cite{PhysRevB.83.205101}. Recently, a family of compounds (Nb/Ta)(P/As) has been experimentally confirmed as WSMs with the remarkable properties such as Fermi arc and chiral anomaly observed\cite{PhysRevX.5.011029,xu2015discoveryTaAs,huang2015weyl,PhysRevX.5.031013DingHongTaAs,NatPhysDingHongTaAs,yang2015weyl,xu2015discoveryNbAs,xu2015experimental,xu2016observation,liu2016evolution,PhysRevX.5.031023,zhang2016signatures,arnold2016negative,hu2015pi,borrmann2015extremely}. The node-line semimetals have also attracted much attention recently, in which the Dirac points form a continuous line or closed ring in the momentum space\cite{PhysRevB.84.235126,PhysRevB.92.205310,weng2016topological,PhysRevLett.117.016602}.

The 112 type compounds AMnBi$_2$ (A= alkaline earth/rare earth metals)\cite{PhysRevB.85.041101,he2012giant,feng2014strong,PhysRevB.87.245104,PhysRevLett.107.126402,PhysRevB.84.220401,PhysRevB.84.064428,PhysRevB.90.075120,PhysRevLett.113.156602,PhysRevB.90.035133,zhang2016interplay,Petrovic2016,wang2016large,masuda2016quantum,PhysRevB.90.075109,borisenko2015time,PhysRevB.94.165161} and A$^{\prime}$MnSb$_2$ (A$^{\prime}$= alkaline earth metals)\cite{0953-8984-26-4-042201,liu2017discovery,liu2016nearly,PhysRevB.95.045128} are known as members of Dirac materials. Interesting transport properties such as quantum oscillations, large unsaturated linear magnetoresistance (MR), nontrivial Berry phase, and coherent interlayer conductivity have been observed in these materials. The Bi/Sb square net is a common feature in these materials and has been considered as the platform of Dirac fermions. Compared to A$^{\prime}$MnSb$_2$, the stronger spin-orbital coupling (SOC) effect from heavy Bi atoms in AMnBi$_2$ always open gaps at Dirac points, which results in larger effective masses of Dirac fermions\cite{liu2016nearly}. Therefore, A$^{\prime}$MnSb$_2$ is a more suitable platform to realize massless Dirac fermions. On the other hand, the introduction of rare earth elements into the place of A$^{\prime}$ in A$^{\prime}$MnSb$_2$ is a guide to design new materials\cite{wollesen1996ternary}.

In order to explore new materials and search for massless Dirac fermions, we synthesized the single crystals of YbMnSb$_2$ and studied the transport properties. Similar to other AMnBi$_2$ and A$^{\prime}$MnSb$_2$, YbMnSb$_2$ is shown as a metal with possible antiferromagnetic transition at $T_N$=345 K. Clear SdH and dHvA type oscillations have been observed at low temperature and high field. The extracted values of Berry phase from quantum oscillations are both nontrivial, indicating that YbMnSb$_2$ may host Dirac fermions. The quasi-2D characteristic of Fermi surface and coherent interlayer conductivity are revealed by angular-dependent SdH oscillation frequency and interlayer MR, respectively. Nonlinear Hall resistivity suggests the coexistence of electron and hole carriers in YbMnSb$_2$.

\section{Experimental and crystal structure}

\begin{figure*}[htbp]
\centering
\includegraphics[width=\textwidth]{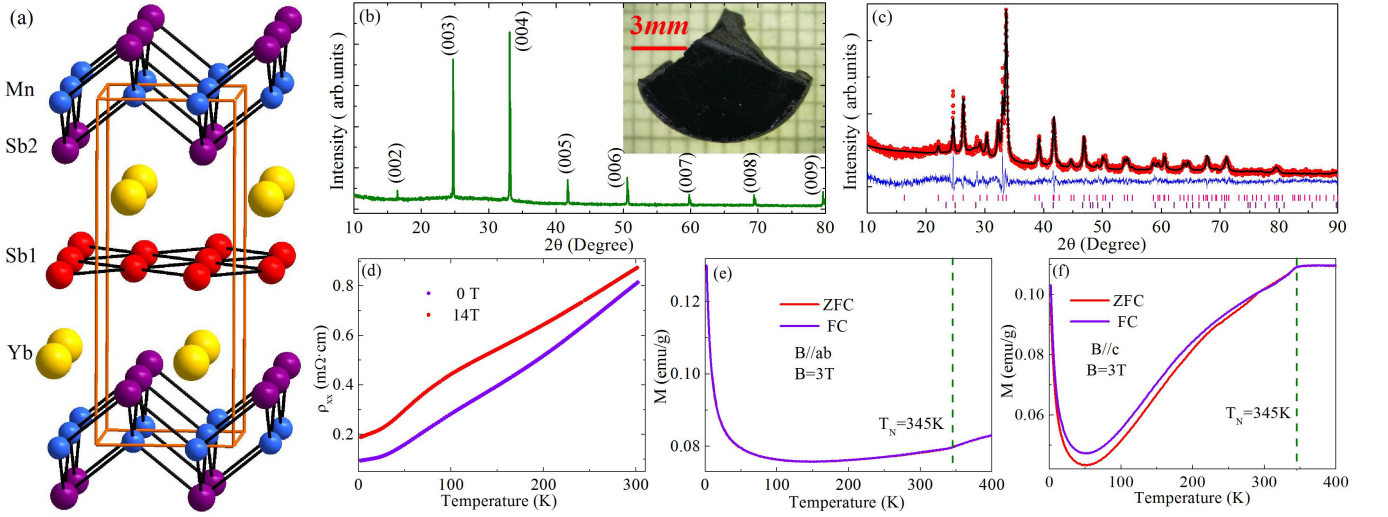}
\caption{(a) The crystal structure of YbMnSb$_2$. (b) XRD of a YbMnSb$_2$ single crystal. Inset shows a typical grown crystal with a scale of 1 cm. (c) Powder XRD pattern with refinement. The experiment data is given by red circle. The black solid line indicates the fit. The difference is plotted in blue. The positions of Bragg peaks of YbMnSb$_2$ and Sb are denoted by pink and purple vertical lines, respectively. (d) Temperature dependence of the in-plane resistivity $\rho_{xx}$ at 0 T and 14 T. (e), (f) Temperature dependence of magnetization with the magnetic field along \emph{ab} plane and \emph{c} axis, respectively.}
\end{figure*}

Large single crystals of YbMnSb$_2$ were grown from antimony flux. The starting elements, Yb, Mn and excess Sb, were placed into a crucible and sealed in a quartz tube with a ratio of Yb: Mn: Sb=1:1:4. The quartz tube was heated to 1050$^0$C in 60 h and then cooled to 690$^0$C at a rate of 1$^0$C/h, at which the excess Sb flux was separated from the crystals by centrifugation. The obtained crystals are as large as 1 cm in its one dimension. The atomic proportion were checked using energy dispersive x-ray spectroscopy (EDS, Oxford X-Max 50). The patterns of single crystal and powder x-ray diffraction (XRD) were collected from a Bruker D8 Advance x-ray diffractometer using Cu K$_{\alpha}$ radiation. TOPAS-4.2 was employed for the refinement. Resistivity measurements were performed on a Quantum Design physical property measurement system (QD PPMS-14T). The magnetic properties were measured with the vibrating sample magnetometer (VSM) option of PPMS.

The powder XRD pattern of YbMnSb$_2$ can be indexed with either $Pnma$ (same as CaMnSb$_2$) or $P4/nmm$ (same as YbMnBi$_2$) space group. The R$_{wp}$ and $\chi^2$ in both structures are extremely similar, and the structure of YbMnSb$_2$ can't be determined only with the refinement of powder XRD patterns. Considering the recent related study on YbMnSb$_2$, where the structure of YbMnSb$_2$ has been indexed to $Pnma$ space group with the single crystal x-ray diffractometer measurement and analysis\cite{Kealhofer2017YbMnSb2}, the crystal structure of YbMnSb$_2$ is illustrated in Fig. 1(a). In the structure of YbMnSb$_2$, each Mn atom is surrounded by four Sb atoms, forming the MnSb$_4$ tetrahedra. YbMnSb$_2$ also possesses a Sb square net (highlighted by red atoms in Fig. 1(a)), which is similar to AMnBi$_2$ and A$^{\prime}$MnSb$_2$. Figure 1(b) shows the single crystal XRD pattern with (00\emph{l}) reflections, indicating that the surface of crystal is \emph{ab} plane. The powder diffraction pattern could be well indexed in the $P4/nmm$ space group (Fig. 1(c)). The determined lattice parameters are $a=b={4.324(0)\AA}$ and $c={10.839(8)\AA}$.

\section{Results and discussion}

\begin{figure*}[htbp]
\centering
\includegraphics[width=\textwidth]{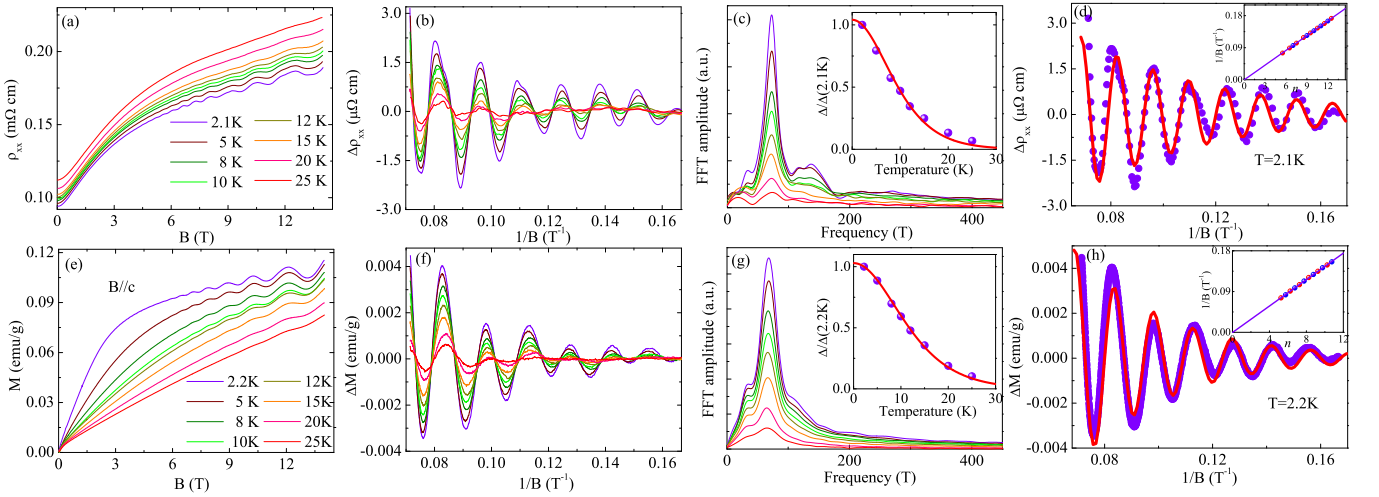}
\caption{Quantum oscillations in YbMnSb$_2$. (a), (e) Magnetic field dependence of resistivity and magnetization at various temperatures. (b), (f) The amplitude of SdH and dHvA oscillations plotted as a function of 1/B. (c), (g) FFT spectra of the oscillations. The insets are the temperature dependence of relative FFT amplitude of the main oscillation frequency. The red solid lines are the fits based on LK formula. (d), (h) The LK fits (red lines) of SdH and dHvA oscillations at 2.1 K and 2.2 K, respectively. The insets are the corresponding Landau level (LL) index fan diagrams. For SdH oscillation, the valleys (red circles) and peaks (blue circles) are assigned as half integer and integer LL indices, respectively. For dHvA oscillation, the valleys (red circles) and peaks (blue circles) are assigned as $n+\frac{1}{4}$ and $n-\frac{1}{4}$ LL indices, respectively.}
\end{figure*}

Figure 1(d) presents the temperature dependence of in-plane resistivity $\rho_{xx}$. $\rho_{xx}$ exhibits a metallic behavior, and the application of magnetic fields enhances the value of $\rho_{xx}$. Figures. 1(e) and 1(f) present the temperature dependent magnetization measured by applying magnetic field along \emph{ab} plane and \emph{c} axis. Below 345 K, the decrease of M(T) for \emph{B//c} and \emph{B//ab} indicates the emergence of possible long range antiferromagnetic order. M(T) reduces sharply for \emph{B//c} while slightly for \emph{B//ab}, suggesting that \emph{c} axis is the magnetic easy axis (demagnetizing effects have been considered). At low temperatures, significant magnetization upturns are observed for both \emph{B//ab} and \emph{B//c}, which may originate from the development of a ferromagnetic component due to canted antiferromagnetic moments\cite{liu2017discovery}.

Figure 2 shows the quantum oscillations in YbMnSb$_2$. As shown in Figs. 2(a) and 2(e), clear SdH and dHvA type oscillations can be observed at low temperature and high field. With temperature increased, the oscillations become weaker gradually. The oscillatory components of resistivity and magnetization are obtained by subtracting a smooth background. Figures 2(b) and 2(f) plot the oscillation amplitudes of resistivity ($\Delta\rho_{xx}=\rho_{xx}-<\rho_{xx}>$) and magnetization ($\Delta M=M-<M>$) against 1/B, respectively. A nearly single frequency ($F_{SdH}=73.0T$, $F_{dHvA}=73.4T$) has been derived from the fast Fourier transformation (FFT) spectra for both SdH (Fig. 2(c)) and dHvA (Fig. 2(g)) oscillations. According to the Onsager relation $F=(\phi_0/2\pi^2)A_F=(\hbar/2\pi e)A_F$, the frequency $F$ is proportional to the cross sectional area $A_F$ of FS normal to the field. The frequencies $F_{SdH}$ and $F_{dHvA}$ in YbMnSb$_2$ are smaller than those in (Ca/Sr/Yb)MnBi$_2$\cite{PhysRevB.85.041101,he2012giant,PhysRevB.84.220401,PhysRevB.94.165161} but greater than those in (Ca/Sr/Ba)MnSb$_2$\cite{liu2017discovery,liu2016nearly,PhysRevB.95.045128}, which indicates that YbMnSb$_2$ possesses a moderate FS area in the Mn-based 112 family. The relationship between structure type/magnetic structure/lattice parameters and FS area demands further study.

In SdH and dHvA oscillations, the amplitude can be described by Lifshitz-Kosevich (LK) formula\cite{shoenberg1984magnetic},
\begin{equation}\label{equ1}
\centering
\Delta\rho\propto\frac{\lambda T}{sinh(\lambda T)}e^{-\lambda T_D}cos[2\pi(\frac{F}{B}-\frac{1}{2}+\beta+\delta)]
\end{equation}

\begin{equation}\label{equ2}
\centering
\Delta M \propto -B^{1/2}\frac{\lambda T}{sinh(\lambda T)}e^{-\lambda T_D}sin[2\pi(\frac{F}{B}-\frac{1}{2}+\beta+\delta)]
\end{equation}
where $\lambda= (2\pi^2k_{B}m^*)/(\hbar eB)$. $2\pi \beta$ is the Berry phase, and $T_D$ is Dingle temperature. $\delta$ is a phase shift. $\delta=0$ and $\pm1/8$ for the 2D and 3D systems, respectively. The thermal factor $R_T=(\lambda T)/sinh(\lambda T)$ in L-K formula has been employed to describe the temperature dependence of FFT amplitude. As shown in the insets of Figs. 2(c) and 2(g), the temperature dependence of relative FFT amplitude can be well fitted, and the extracted effective masses are $m^*_{SdH}=0.13m_e$ and $m^*_{dHvA}=0.11m_e$. Small effective mass is usually a characteristic of Dirac fermion. The effective masses in YbMnSb$_2$ are comparable with that of Sr$_{1-y}$Mn$_{1-z}$Sb$_2$ (0.14$m_e$)\cite{liu2017discovery} but slightly larger than those of CaMnSb$_2$ (0.05-0.06$m_e$)\cite{PhysRevB.95.045128} and BaMnSb$_2$ (0.052-0.058$m_e$)\cite{liu2016nearly}.

\begin{table*}
  \centering
  \caption{Parameters derived from SdH and dHvA oscillations. $F$, oscillation frequency; $A_F$, cross sectional area of FS normal to the field; $k_F$, Fermi vector by assuming a circular cross section; $T_D$, Dingle temperature; $m^*$, effective mass; $\tau$, quantum lifetime; $\mu_q$, quantum mobility; $v_F$, Fermi velocity; $E_F$, Fermi energy; $2\pi\beta$, Berry phase.}
  \label{oscillations}
  \begin{tabular}{cccccccccccc}
    \hline\hline
           & $F$ (T) & $A_F$ ({\AA}$^{-2}$) & $k_F$ ({\AA}$^{-1}$) & $T_D$ (K) & $m^*/m_e$ & $\tau$ (10$^{-13}$s) & \makecell[c]{$\mu_q$ \\ (m$^2$/Vs)} & $v_F$ (m/s) & $E_F$ (eV) & \makecell[c]{$2\pi\beta$ \\ ($\delta=0$) \\ (LK fits)} & \makecell[c]{$2\pi\beta$ \\ ($\delta=0$) \\ (LL fan diagrams)} \\
    \hline
     SdH  & 73 & 0.007 & 0.047 & 10.1 & 0.13 & 1.2 & 0.16 & 4.1$\times$10$^5$ & 0.126 & 0.972$\pi$ & 1.056$\pi$ \\
    dHvA & 73.4 & 0.007 & 0.047 & 18.5 & 0.11 & 0.7 & 0.11 & 5.0$\times$10$^5$ & 0.157 & 1.007$\pi$ & 1.057$\pi$ \\
    \hline\hline
  \end{tabular}
\end{table*}

Berry phase $2\pi\beta$ can be obtained from the analyses of quantum oscillations. A nontrivial $\pi$ Berry phase is expected for the relativistic Dirac fermions. There are two methods to extract the Berry phase. One is to map the Landau level (LL) index fan diagrams. The LL index \emph{n} is related to 1/B by Lifshitz-Onsager quantization rule $A_F(\hbar/2\pi eB)=n+1/2-\beta-\delta$, the Berry phase can be extracted from the intercept of the linear extrapolation. Another one is to fit the field dependence of oscillation amplitude with LK formula directly. In order to guarantee the reliability of the obtained Berry phase, we extracted the Berry phase from SdH and dHvA type oscillations using both methods (Figs. 2(d), 2(h) and the insets). As shown in the Table \ref{oscillations}, all the obtained values of Berry phase are very close to nontrivial value $\pi$, which indicates the possible existence of Dirac fermions in YbMnSb$_2$.

The LK fits also yield the Dingle temperatures, which relate to the quantum lifetime and quantum mobility by $\tau_q=\hbar/(2\pi k_B T_D)$ and $\mu_q=e\tau_q/m^*$. As shown in Table \ref{oscillations}, the obtained quantum mobilities are $\mu_q (SdH)=0.16 m^2/Vs$ and $\mu_q (dHvA)=0.11 m^2/Vs$, both of which are comparable with that of BaMnSb$_2$ ($0.128 m^2/Vs$)\cite{liu2016nearly} but much higher than those of SrMnSb$_2$ ($0.057 m^2/Vs$)\cite{liu2017discovery} and SrMnBi$_2$ ($0.025 m^2/Vs$)\cite{PhysRevLett.107.126402}. The Fermi velocity  $v_F=\hbar k_F/m^*$ and Fermi energy $E_F=m^* v^2_F$ derived from quantum oscillations have also been listed in Table \ref{oscillations}.

\begin{figure}[htbp]
\centering
\includegraphics[width=0.48\textwidth]{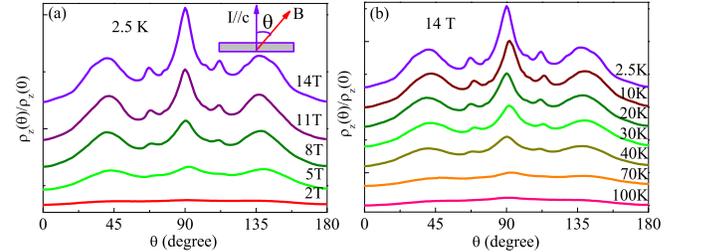}
\caption{(a) The normalized interlayer resistivity $\rho_z(\theta)/\rho_z(0)$ at T=2.5 K and different magnetic fields. The inset shows the definition of $\theta$. (b) The normalized interlayer resistivity $\rho_z(\theta)/\rho_z(0)$ at B=14 T and different temperatures. In (a) and (b), the data are shifted along the vertical axis for clarity.}
\end{figure}

In layered metals, an interesting topic is to study the interlayer transport properties. The motion of carriers between the layers may be coherent or incoherent\cite{PhysRevLett.81.4492,Valla2002Coherence}. For the incoherent interlayer transport, FS is only defined within the plane of layers since the scattering rate in the layer is much larger than the interlayer hopping integral\cite{PhysRevB.65.180506}. The interlayer conductivity is proportional to the tunneling rate between layers\cite{PhysRevLett.81.4492,PhysRevLett.109.147005}. However, in the case of coherent interlayer transport, the interlayer conductivity is determined by the carriers scattering rate and the electronic group velocity perpendicular to the layers\cite{PhysRevLett.81.4492,Valla2002Coherence}. As presented in Figs. 3(a) and 3(b), the interlayer resistivity $\rho_z(\theta)$ exhibits a sharp peak when $\theta=90^0$, at which the magnetic field is rotated into the \emph{ab} plane. The intensity of the peak decreases gradually with increasing temperature or decreasing field. The peak is attributed to the small closed orbits\cite{PhysRevB.57.1336} or the self-crossing orbits\cite{PhysRevB.60.11207,kartsovnik2004high} on the side of warped quasi-2D FS. The angular dependence of interlayer MR reveals the coherent interlayer transport in YbMnSb$_2$. There are several additional peaks around $\theta=90^0$, which are called Yamaji peaks\cite{yamaji1989angle}. Yamaji peaks originate from the geometrical effect of quasi-2D FS\cite{yagi1990semiclassical,PhysRevLett.105.246403,PhysRevLett.113.156602}, and the peak positions are independent of the magnetic field strength.

\begin{figure}[htbp]
\centering
\includegraphics[width=0.48\textwidth]{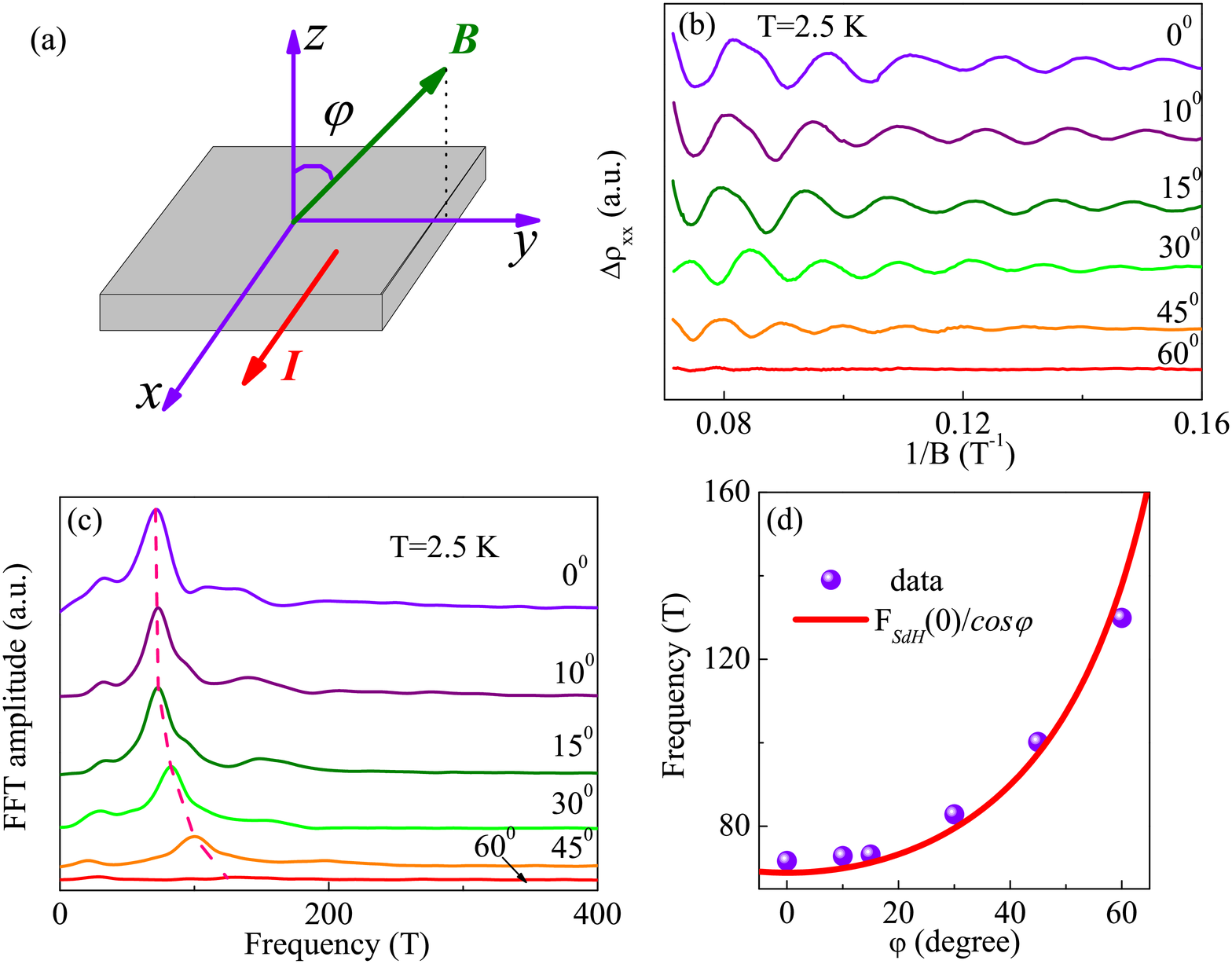}
\caption{(a) The definition of $\varphi$. (b) and (c) are the oscillatory components and FFT spectra of SdH oscillation under different $\varphi$, respectively. The temperature is fixed at 2.5 K. In (b) and (c), the data are shifted along the vertical axis for clarity. (d) The angle dependence of the main SdH oscillation frequency determined from FFT spectra. The red solid line is the fit to $F(\varphi)=F(0)/cos\varphi$.}
\end{figure}

To get more information about the FS of YbMnSb$_2$, we perform the measurements of angular-dependent transverse MR. As shown in Fig. 4(a), the current \textbf{\emph{I}} flows in \emph{ab} plane and the magnetic field \textbf{\emph{B}} rotates in the plane perpendicular to the current. $\varphi$ is the angle between \emph{c} axis and the field. Figures. 4(b) and 4(c) show the oscillatory component and frequency of SdH oscillation under different angle $\varphi$. As the magnetic field is rotated from $\varphi=0^0$ to $\varphi=60^0$, the SdH oscillation pattern and FFT spectra evolve systematically. The main oscillation frequency follows $F(\varphi)=F(0)/cos\varphi$ and can be well fitted as shown in Fig. 4(d), which suggest that the FS responsible for SdH oscillation is quasi-2D.

\begin{figure}[htbp]
\centering
\includegraphics[width=0.48\textwidth]{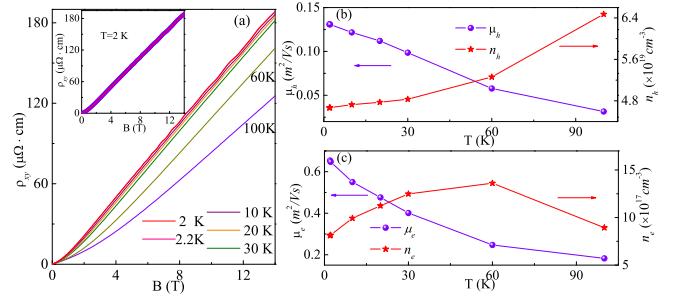}
\caption{(a) The magnetic field dependence of Hall resistivity of YbMnSb$_2$ at various temperatures. The inset shows the two-band model fitting (red solid line) of the Hall data at 2 K. (b) The temperature dependence of mobility ($\mu_h$) and concentration ($n_h$) of holes. (c) The temperature dependence of mobility ($\mu_e$) and concentration ($n_e$) of electrons.}
\end{figure}

Figure 5(a) displays the magnetic field dependence of the Hall resistivity $\rho_{xy}$ at different temperatures. Nonlinear behavior of $\rho_{xy}$ indicates the coexistence of electron and hole in YbMnSb$_2$. Thus, semiclassical two-band model is employed to describe the Hall resistivity:
\begin{equation}\label{equ3}
\rho_{xy}=\frac{B}{e}\frac{(n_h \mu_h^2-n_e \mu_e^2)+(n_h-n_e)(\mu_h \mu_e)^2 B^2}{(n_h \mu_h+n_e \mu_e)^2+(n_h-n_e)^2 (\mu_h \mu_e)^2 B^2}
\end{equation}
where $n_e(n_h)$ and $\mu_e(\mu_h)$ correspond to electron (hole) concentration and electron (hole) mobility, respectively. The fitted curve (red solid line) is consistent with the experimental curve (violet dots) as shown in the inset of Fig. 5(a). Figures 5(b) and 5(c) show the temperature dependence of the carrier concentration and mobility of hole and electron, respectively. The carrier concentrations increase with increasing temperature. But for the concentration of electron, there is a slight drop from 60 K to 100 K. The concentrations at 2 K are $n_h=4.67\times10^{19} cm^{-3}$ and $n_e=8.10\times10^{17} cm^{-3}$, respectively. With the temperature increased, both the mobilities of hole and electron reduced. The mobilities at 2 K are $\mu_h=1310 cm^2V^{-1}s^{-1}$ and $\mu_e=6538 cm^2V^{-1}s^{-1}$, respectively.

\section{Summary}

In summary, we have grown high quality single crystals of a new compound YbMnSb$_2$ and investigated the transport properties. YbMnSb$_2$ exhibits metallic behavior and shows a possible antiferromagnetic transition at $T_N$=345 K. Both SdH oscillation and dHvA type oscillation are observed in YbMnSb$_2$. The analysis of quantum oscillations yields nontrivial $\pi$ Berry phase and small effective masses, which are usually the characteristics of Dirac fermions. Coherent interlayer conductivity has been revealed by the angular-dependent interlayer MR. The angular dependence of the main SdH oscillation frequency suggests the quasi-2D feature of corresponding FS. YbMnSb$_2$ contains two types of carriers, and the carrier concentrations and mobilities are obtained by fitting the Hall data. Our work indicates that YbMnSb$_2$ is a new candidate of Dirac semimetal with coherent interlayer transport and quasi-2D FS.

\emph{Note added.} When the paper is being finalized, we notice one related work on YbMnSb$_2$\cite{Kealhofer2017YbMnSb2}.

\section{Acknowledgments}

This work is supported by the National Natural Science Foundation of China (No.11574391), the Fundamental Research Funds for the Central Universities, and the Research Funds of Renmin University of China (No. 14XNLQ07).

\bibliography{Bibtex}
\end{document}